\documentclass[sigconf]{acmart}
\usepackage[utf8]{inputenc}
\usepackage{booktabs}
\usepackage{multirow}
\usepackage{ctable}
\usepackage{tabularx}
\usepackage{graphicx}
\usepackage{array}
\usepackage[all]{nowidow}
\usepackage{balance}
\usepackage{xspace}
\usepackage{url}
\newcommand{\omt}[1]{}
\newcommand{\firstmention}[1]{{\bf #1}}

\newcommand{\myparagraph}[1]{\vspace{0.3\baselineskip}\noindent{\textbf{#1}}.~}

\newcommand{\query}{q}
\newcommand{\doc}{d}

\newcommand{\curDoc}{\doc_{cur}}
\newcommand{\newDoc}{\doc_{next}}
\newcommand{\tfVec}[1]{\vec{#1}^{T}}
\newcommand{\tfVecParm}[2]{\vec{#1}^{T}_{#2}}
\newcommand{\planted}{planted\xspace}

\newcommand{\wVec}[1]{\vec{#1}^{W}}
\newcommand{\wVecParm}[2]{\vec{#1}^{W}_{#2}}

\newcommand{\psg}{g}
\newcommand{\genRank}{\pi}
\newcommand{\psgNext}{\psg_{follow}}
\newcommand{\psgPrev}{\psg_{prec}}
\newcommand{\curRank}{\genRank_{cur}}
\newcommand{\nextRank}{\genRank_{next}}
\newcommand{\psgSrc}{\psg_{src}}
\newcommand{\psgTarget}{\psg_{target}}
\newcommand{\qryTermSrc}{QryTermSrc\xspace}
\newcommand{\qryTermTarget}{QryTermTarget\xspace}
\newcommand{\stf}{T}
\newcommand{\sWV}{W}
\newcommand{\simSrcTopTF}{SimSrcTop(\stf)\xspace}
\newcommand{\simSrcTopWtV}{SimSrcTop(\sWV)\xspace}
\newcommand{\simTargetTopTF}{SimTargetTop(\stf)\xspace}
\newcommand{\simTargetTopWtV}{SimTargetTop(\sWV)\xspace}
\newcommand{\simSrcPrevTopTF}{SimSrcPrevTop(\stf)\xspace}
\newcommand{\simSrcPrevTopWtV}{SimSrcPrevTop(\sWV)\xspace}
\newcommand{\simTargetPrevTopTF}{SimTargetPrevTop(\stf)\xspace}

\newcommand{\simTargetPrevTopWtV}{SimTargetPrevTop(\sWV)\xspace}
\newcommand{\simSrcTargetWtV}{SimSrcTarget(\sWV)\xspace}
\newcommand{\simSrcPrevWtV}{SimSrcPrecPsg(\sWV)\xspace}
\newcommand{\simSrcNextWtV}{SimSrcFollowPsg(\sWV)\xspace}
\newcommand{\simTargetPrevWtV}{SimTargetPrecPsg(\sWV)\xspace}
\newcommand{\simTargetNextWtV}{SimTargetFollowPsg(\sWV)\xspace}
\newcommand{\rankPromoteLabel}{r}
\newcommand{\coherenceLabel}{c}

\newcommand{\psgSet}[1]{G(#1)}
\newcommand{\psgPool}{G_{pool}}

\newcommand{\set}[1]{\{#1\}}
\newcommand{\definedas}{\stackrel{def}{=}}

\begin{document}
\title{Ranking-Incentivized Quality Preserving Content Modification}
\author{Gregory Goren}
  \email{gregory.goren@campus.technion.ac.il}
\affiliation{%
  \institution{Technion}
}

\author{Oren Kurland}
\email{kurland@technion.ac.il}
\affiliation{%
  \institution{Technion}
}

\author{Moshe Tennenholtz}
\email{moshet@ie.technion.ac.il}
\affiliation{%
  \institution{Technion}
}

\author{Fiana Raiber}
\email{fiana@verizonmedia.com}
\affiliation{%
  \institution{Yahoo Research}
}

\begin{abstract}
The Web is a canonical example of a competitive retrieval setting
where many documents' authors consistently modify their documents to
promote them in rankings. We present an automatic method for
quality-preserving modification of document content --- i.e.,
maintaining content quality --- so that the document is ranked higher
for a query by a non-disclosed ranking function whose rankings can be
observed. The method replaces a passage in the document with some
other passage. To select the two passages, we use a learning-to-rank
approach with a bi-objective optimization criterion: rank promotion
and content-quality maintenance. We used the approach as a bot in content-based ranking competitions. Analysis of the
competitions demonstrates the merits of our approach with respect to
human content modifications in terms of rank promotion,
content-quality maintenance and relevance.
\end{abstract}

\maketitle

\section{Background and Motivation}
\label{sec:intro}
Several research communities nurture work on adversarial attacks
on algorithms. The motivation is to push the sate-of-the-art by
identifying model and algorithmic weaknesses. The ``attacked''
algorithms are often used in real-life systems (e.g., face
recognition~\cite{Dong+al:19a}). Exposing their vulnerabilities is considered an accelerator for innovation more than
a threat.

A classic example is the crypto community. Throughout the decades,
publications of successful attacks on crypto mechanisms helped to push
forward improved mechanisms \cite{Boneh:99a}. Additional examples are the
machine learning, natural language
processing and vision communities. There has recently
been much work on devising adversarial attacks on machine learning
algorithms --- specifically neural networks --- that span different
tasks: general machine learning challenges
\cite{Szegedy+al:14a,Huang+al:17a,Papernot+al:17a}, reading
comprehension \cite{Jia+Liang:17a}, speaker identification
\cite{Kreuk+al:18a}, object detection \cite{Xie+al:17a}, face
recognition \cite{Dong+al:19a}, and more. This line of work has
driven forward the development of algorithms which are more robust to
adversarial examples; e.g., Jia et al. \cite{Jia+al:19a}, He et al. \cite{He+al:17a}, Zhang et
al. \cite{Zhang+al:19a}.

The Web search echo-system is, perhaps, the largest-scale adversarial
setting in which algorithms, specifically search methods,
operate. That is, many document authors consistently modify their
documents to have them more highly ranked for specific queries.  This
practice is often referred to as search engine optimization (SEO)~\cite{Gyongyi+Molina:05a}.  The incentive is quite clear: high ranks
translate to high utility as most of the attention --- and therefore
engagement --- of search engine users is focused on the documents most
highly ranked \cite{Joachims+al:05a}. Some SEO techniques are
considered ``illegitimate'' (a.k.a., black hat~\cite{Gyongyi+Molina:05a}) as they are unethical and hurt the echo-system (e.g., search effectiveness and/or user experience); spamming is a prominent
example. Other techniques are considered completely ``legitimate''
(a.k.a., white hat \cite{Gyongyi+Molina:05a}) as they are intended to
improve documents' representations with respect to queries to which
they pertain. Thus, in the ``ranking games'' that take place on the
Web~\cite{Tennenholtz+Kurland:19a}, the documents' authors are
``players'' or ``adversaries'' whose adversarial actions can be
``legitimate'' (white hat) or ``illegitimate'' (black hat); the
search engine's ranking function is the mediator.

Despite its importance as a large scale and highly evolved adversarial
setting, and despite the research attention paid to adversarial
attacks and defenses in other research communities as mentioned above,
the adversarial effects in the Web search echo-system have attracted
relatively little research attention \cite{AIRWeb,Castillo+Davison:10a}.
An important reason for this reality is that developing black hat SEO
techniques is unethical, and can hurt the search echo-system in the
short term. Yet, it can also potentially help to find vulnerabilities
of search functions which is important for improving them in the long
run. We subscribe to the stand that, despite the potential long term
merits, such type of work should see no place in
scientific publications.

Still, an important question is why there is not a
much larger body of work on addressing adversarial effects. The vast majority of such work
has been on spam identification
\cite{AIRWeb,Castillo+Davison:10a}. One part of the answer seems to be
that many of the adversarial effects are due to, or involve, the dynamic nature of the Web --- changes of pages,
ranking function, indexing cycles, etc. For example, the fact that some authors
are incentivized to compete for improved ranking yields, in 
white hat settings, negative
impact on the search echo-system; specifically, degrading topical
diversity in the corpus~\cite{Basat+al:17a,Raifer+al:17a}\footnote{Similar dynamics in recommendation systems was also studied \cite{BenPorat+Tennenholtz:18a}.}.
Studying such dynamics in terms of rankings seems to require access to query
logs of large-scale commercial Web search engines. In other words, it
is very difficult to impossible to replicate the dynamic search
setting on the Web. In contrast, for example, devising and evaluating content-based
spam classifiers can be done using a static snapshot of documents and
regardless of rankings.

We claim that the state-of-affairs just described, in terms of the
limited volume and scope of work on adversarial Web search, is
nowadays being challenged to a major extent and there is a call to
action. In the last few years there has been a dramatic change in
terms of the potential ability to generate large scale and high
quality black hat content effects. For example, advanced language modeling
techniques such as BERT \cite{Delvin+al:18a}, GPT2
\cite{Radford+al:18a} and XLNet \cite{Yang+al:19a} can be used to
automatically generate fake content at scale
\cite{Zellers+al:19a}. While fake content could be considered spam, it
is way more difficult to identify than classic spam, especially when
generated using the above mentioned state-of-the-art techniques
\cite{Zellers+al:19a}; as a case in point, fake content could still be of high
quality.  Another example for a
modern threat on search engines is that for neural-based retrieval
methods \cite{Mitra+al:18a}. It could be the case that adversarial
attacks on neural methods used for vision and NLP applications as
mentioned above will soon be translated to attacks on ranking methods.



One of the important implications of the reality just described is
that Web document authors who will not use automatic tools for content
creation and manipulation, specifically for automatic white-hat SEO
which will help their documents to be highly ranked when they deserve
to, will not be competitive in the adversarial search setting. This
implication, together with the fact that creating benchmarks that will
allow to research dynamic and adversarial search settings remains
highly difficult to impossible to achieve, motivate our work.  That
is, given the increased ability to hurt the search echo-system, we
strive to counter balance it by developing legitimate, white hat,
tools that can (i) help Web authors in the competitive search setting
without hurting the search echo-system, and (ii) be used for creating
publicly available benchmarks that will allow to study competitive and
adversarial dynamic search settings.

\section{Contribution}
We present the first, to the best of our knowledge,
attempt to devise an automatic method of ``legitimate'', white hat, content modification of documents. The goal of the modification is promotion in rankings
induced by a non-disclosed relevance ranking function for a given
query. The method can observe past rankings for the query
which are the only signals about the ranking function. By
``legitimate'' modification we mean a change that maintains the
document's content quality in contrast to black hat SEO
\cite{Gyongyi+Molina:05a}. 

Our method replaces a short passage of the
document with another short passage from a candidates pool.
We cast this passage replacement task as a learning-to-rank (LTR) \cite{Liu:11a} problem over passage pairs:
a passage in the document and a
candidate passage for replacement. The optimization goal for
training the LTR function is bi-objective: rank promotion and
content-quality maintenance which we address via presumed coherence maintenance.
The highest
ranked passage pair is selected for the replacement.

We evaluated our approach by using it as a bot
in content-based ranking competitions we organized between students\footnote{The dataset is at: \url{https://github.com/asrcompetition/content_modification_dataset}; the code is at \url{https://github.com/asrcompetition/content_modification_code}.}.
The competitions were approved by international and institutional ethics committees.
In the competitions, the bot produced documents which were promoted in
rankings to a larger extent than the students' documents. Furthermore,
the bot's content modifications did not hurt relevance in
contrast to students' modifications of their documents, and maintained
content quality to a large extent.

Hence, although simple, our approach constitutes a first {\em
  scientific} proof of concept for the feasibility of manipulating
document content for rank promotion in search engines while
maintaining the document quality. It is important to keep in mind that
our focus in this paper is on the basic task of selecting passages of
the document, and passages from some given pool, to perform the
replacement. Creating the pool of candidates for replacement is an
important task at its own right which we discuss as a future work;
e.g., leveraging the recent significant progress in automatic language
generation capabilities mentioned above. For the proof
of concept in this paper, we simply used a pool of passages extracted
from other documents which were highly ranked for the given query in
past rankings. The motivation for this practice is based on some recent observations about SEO strategies
employed by publishers \cite{Raifer+al:17a}; namely, that they tend to mimic documents highly ranked in the past. Obviously, this is not
a practical solution for pool generation due to copyright issues, but rather a
means to our end of evaluating our proposed learning-based approach for passage
replacement. In addition to devising methods for creating a pool of candidate passages, moving towards a practical application of our approach
will call for introducing modifications which are not
content-based. While content-based features are extremely important in
Web ranking functions \cite{Liu:11a}, there are other types of important features.


The line of research we pursue is important not only for
document authors so as to ``keep up'' with the ranking game in a legitimate manner, but also for those who devise ranking functions in
adversarial retrieval settings. That is, having document modification
methods will allow to create a myriad of benchmarks which do not exist today for studying dynamic retrieval settings, even if in our case these are white hat.

\section{Related Work}
\label{sec:rel}
There is a body of work on identifying/fighting black hat SEO; specifically,
spam \cite{AIRWeb,Castillo+Davison:10a}. Our 
approach is essentially a content-based white hat SEO method intended to rank-promote
{\em legitimate} documents via {\em legitimate}
content modifications. We are not aware of past work on devising such automatic content modification procedures. 

Our approach might seem at first glance conceptually similar to
the black hat {\em stitching} technique \cite{Gyongyi+Molina:05a}:
authors of low-quality Web pages manually ``glue'' to their
documents unrelated phrases from other documents. In contrast, our approach operates on descent quality
documents and is optimized to maintain document quality.

Our approach can conceptually be viewed as ranking-incentivized
paraphrasing: modifying the document to promote
it in rankings, but keeping content quality and having the content remain faithful to the original content. Past work on paraphrasing (e.g., \cite{Androutsopoulos+Malakasiotis:10a}) does not include methods
intended to promote documents in ranking.

We use simple estimates (e.g., lexical and Word2Vec similarities) to
measure the extent to which content coherence is maintained given the passage
replacement. More evolved estimates can be used to this end \cite{graesser2004coh,lapata2005automatic,pitler2008revisiting,lin2011automatically,li2016neural}. Furthermore, one could modify the document using text-generation approaches that account for coherence \cite{kiddon2016globally,ji2016latent,guo2018long}, which we leave for future work.

\section{Content Modification Approach}
\label{sec:framework}
Suppose that the author of a document with descent content quality would like the document to be highly ranked for a
query $\query$ by a search engine whose ranking function is not
disclosed. More specifically, the author observes the current ranking,
$\curRank$, induced for $\query$, and her goal is to promote her
document, $\curDoc$, in the next induced ranking, $\nextRank$,
assuming that $\curDoc$ was not the highest ranked in $\curRank$. We
present an approach to automatically modifying $\curDoc$'s content to
this end, yielding a document $\newDoc$.\footnote{The ``next ranking''
  is induced after the search engine indexed
  $\newDoc$ instead of $\curDoc$.}

There are three desiderata for the content modification: (i) maximizing the likelihood that $\newDoc$ will be
positioned in $\nextRank$ at a higher rank than $\curDoc$'s current
rank in $\curRank$, (ii) maintaining content quality, 
and (iii) having $\newDoc$ faithful to $\curDoc$ in
terms of the provided information.


In reference to work on passage-based document paraphrasing (e.g.,
\citet{Barzilay+Lee:03a}), we perform the content modification by
replacing one of $\curDoc$'s short passages with another short passage from
a given pool of candidates, $\psgPool$.\footnote{The approach does not depend on the passage markup technique.} The goal is to optimize the replacement with respect to the desiderata mentioned above.
We cast the task as a learning-to-rank challenge \cite{Liu:11a} over pairs of passages, $(\psgSrc,\psgTarget)$,
where $\psgSrc \in \psgSet{\curDoc}$ and $\psgTarget \in \psgPool$;
$\psgSet{\curDoc}$ is the set of $\curDoc$'s passages. The highest ranked passage pair is used for replacement. 

Candidate passages in the pool $\psgPool$ can be created in various
ways; e.g., from scratch using language generation techniques, or by paraphrasing passages in $\curDoc$ or
those in other documents. However, our focus here is on the passage-pair
ranking challenge, and more generally, on the first proof of concept for the content-modification challenge we pursue. Hence, we used a simple approach to create $\psgPool$
following recent findings about strategies employed by
documents' authors to promote their documents in rankings \cite{Raifer+al:17a}: authors tend to mimic content in documents that were highly
ranked in the past for a query of interest. The
merits of this strategy were demonstrated using theoretical and
empirical analysis \cite{Raifer+al:17a}. The simple motivation behind this strategy is
that induced rankings are the only (implicit) signal about the ranking
function, and documents highly ranked are examples for what the
ranking function rewards. Accordingly, here, $\psgPool$ is
the set of all passages in documents ranked higher than $\curDoc$ in
$\curRank$. In practical applications, these passages will not be used directly to avoid copyright issues. As noted above, they can be paraphrased, or the passage pool creation can alternatively rely on automatic passage generation. We leave these challenges for future work.

\subsection{Learning to Replace Passages}
\label{sec:labels}



The ranking function for passage pairs, $(\psgSrc,\psgTarget)$, should be optimized for the
three desiderata described above. Here we focus on the first two ---
rank promotion and maintaining content quality. Since content quality
is a difficult notion to quantify, we set as a goal to not
significantly hurt ``local coherence'' in terms of the passage relations (e.g., semantic similarities) to its surrounding passages.


We do not directly address the 
desideratum of $\newDoc$'s faithfulness to $\curDoc$. Yet, using the coherence-based features suggested below and the fact that $\psgSrc$ and $\psgTarget$ are short passages help to 
keep $\newDoc$ ``semantically similar'' to $\curDoc$.

Our passage-pair ranking function is optimized, simultaneously,
for achieving rank promotion and maintaining local coherence. This is a dual-objective
optimization. Inspired by work on learning Web ranking functions
with a dual objective: relevance and freshness~\cite{Dong+al:10a,Dong+al:10b,Dai+al:11a}, we use labels which are aggregates of rank-promotion and coherence labels. 

Specifically, if $\curDoc$ is a document in the training data, and $\curRank$ is the current ranking it is part of, we
produce a rank-promotion label $\rankPromoteLabel$ with values in
$\set{0, 1, \ldots, 4}$ and a local-coherence maintenance label $\coherenceLabel$ with
values in $\set{0, 1, \ldots, 4}$ for each pair $(\psgSrc,\psgTarget)$
in $ \psgSet{\curDoc} \times \psgPool$.
Details about producing these labels are provided in Section
\ref{sec:eval}. We then produce a single label $l$ ($\in \set
    {0,\ldots,4})$ for $(\psgSrc,\psgTarget)$ by aggregating
    $\rankPromoteLabel$ and $\coherenceLabel$ using the (smoothed) harmonic mean \cite{Dai+al:11a}:
    $l \definedas
    \frac{(1+\beta^2)\rankPromoteLabel
      \coherenceLabel}{\rankPromoteLabel + \beta^2 \coherenceLabel +
      \epsilon}$, where $\beta$ is a free parameter that controls the
    relative weight assigned to the rank-promotion and coherence
    labels, and $\epsilon = 10^{-4}$ is a smoothing parameter.

We can now use these labels that quantify, simultaneously,
rank-promotion and local coherence maintenance in any learning-to-rank approach
\cite{Liu:11a} to learn a ranking function for passage pairs.


\subsubsection{Features for Passage Pairs}
\label{sec:features}
The passage pair $(\psgSrc,\psgTarget)$ is represented by a feature
vector with two types of features: those that target potential rank
promotion as a result of moving from $\curDoc$ to $\newDoc$, and those that target the
change of local coherence as a result of this move. None of the features is based on assuming knowledge of the ranking function for which promotion is sought.

Herein, 
$\tfVec{x}$ denotes the TF.IDF vector representing text $x$ (a document or a passage); $\wVec{x}$ denotes its Word2Vec-based vector representation \cite{Mikolov+al:13a}: we average
the Word2Vec vectors of terms in a passage to yield a passage vector,
and we average the resultant vectors of passages in a document to yield a document
vector. We provide details of training Word2Vec in Appendix \ref{sec:app}. We measure the similarity between two
vectors using the cosine measure.

\myparagraph{Rank-promotion features}

The \qryTermSrc and \qryTermTarget features are the fraction of
occurrences of $\query$'s terms in $\psgSrc$ and $\psgTarget$,
respectively. The assumption is that document retrieval scores assigned by
any retrieval method increase with increased query-terms frequency \cite{Fang+Zhai:05a}. 

The {\simSrcTopTF}, {\simTargetTopTF}, {\simSrcTopWtV} and
{\simTargetTopWtV} features are
$\cos(\tfVecParm{\psg}{src},cent^{T}_{\curRank})$, $\cos(\tfVecParm{\psg}{target},cent^{T}_{\curRank})$,
$\cos(\wVecParm{\psg}{src},cent^{W}_{\curRank})$ and
$\cos(\wVecParm{\psg}{target},cent^{W}_{\curRank})$, respectively; $cent^{T}_{\curRank}$ and
$cent^{W}_{\curRank}$ are the arithmetic mean (centroids) of the $m$ TF.IDF-based
and Word2Vec-based vectors, respectively, that represent the $m$ most
highly ranked documents in the current ranking $\curRank$ that are also ranked
higher than $\curDoc$. We set the maximum value of $m$ to~$3$. As
the ranking function is unknown, similarity to top-retrieved
documents can somewhat attest to increased retrieval score.

The next four features are based on the assumption that similarity of a passage to documents which were highly ranked in the past for the query can attest to improved retrieval score.
Formally, we observe the $p$ past and current rankings induced for $\query$:
$\genRank_{-1}, \genRank_{-2}, \ldots$, $\genRank_{-p}$; $\genRank_{-1}$ is the current ranking $\curRank$ and $\genRank_{-2}$ is the previous ranking; $p$ is a free parameter. Let $\doc_{\genRank_{-i}}$ be the document
most highly ranked in $\genRank_{-i}$; we set $cent^{T}_{\genRank_{past}}
  \definedas \sum_{i=1}^{p} w_i \tfVecParm{\doc}{\genRank_{-i}}$ and
    $cent^{W}_{\genRank_{past}} \definedas
      \sum_{i=1}^{p} w_i \wVecParm{\doc}{\genRank_{-i}}$; $w_i$ is a time-decay-based weight:
$w_i \definedas \frac{\alpha \exp (-\alpha i )  }{\sum_{j=1}^{p} \exp (-\alpha
  j)}$, which is inspired by work on time-based
language models \cite{Li+Croft:03a} with $\alpha=0.01$ \cite{Li+Croft:03a}. Then, the features {\simSrcPrevTopTF}, {\simTargetPrevTopTF}, {\simSrcPrevTopWtV} and
{\simTargetPrevTopWtV} are defined as
$\cos(\tfVecParm{\psg}{src},cent^{T}_{\genRank_{past}})$, $\cos(\tfVecParm{\psg}{target},cent^{T}_{\genRank_{past}})$, $\cos(\wVecParm{\psg}{src},cent^{W}_{\genRank_{past}})$ and
$\cos(\wVecParm{\psg}{target},cent^{W}_{\genRank_{past}})$, respectively.

\myparagraph{Coherence-maintenance features}
The next features, all
based on Word2Vec similarities, address local coherence. The {\simSrcTargetWtV} feature is: $\cos
(\wVecParm{\psg}{src},\wVecParm{\psg}{target})$.

The next four
features are similarities of $\psgSrc$ and $\psgTarget$ with the
context of $\psgSrc$ in $\curDoc$: its preceding and following passages in
$\curDoc$ denoted $\psgPrev$ and $\psgNext$, respectively. Specifically, {\simSrcPrevWtV}, {\simSrcNextWtV},
{\simTargetPrevWtV} and {\simTargetNextWtV} are:
$\cos(\wVecParm{\psg}{src},\wVecParm{\psg}{prec})$,
$\cos(\wVecParm{\psg}{src},\wVecParm{\psg}{follow})$,\\
$\cos(\wVecParm{\psg}{target},\wVecParm{\psg}{prec})$
and $\cos(\wVecParm{\psg}{target},\wVecParm{\psg}{follow})$. If
  $\psgSrc$ is the first passage in $\curDoc$ then we use $\psgNext$
  instead of $\psgPrev$; if $\psgSrc$ is the last passage in
  $\curDoc$ we use $\psgPrev$ instead of $\psgNext$; i.e., in both
  cases, the same feature is used twice.

\section{Evaluation}
\label{sec:eval}
Our document modification approach operates as
follows. First, a ranking for a query is observed. Then, the approach
is applied to modify a given document with the goal that the resulting document
will be ranked higher in the next induced ranking. In real dynamic
settings, other documents can change at the same time thereby
affecting the next ranking. Accordingly, we devise two types of evaluation, online and offline, both based on a dynamic setting.


\subsection{Experimental Setting}
\label{sec:expSet}
To perform the online evaluation, we used our approach as a bot in
live content-based ranking competitions that we organized. The competitions were inspired by
those presented by \citet{Raifer+al:17a} who analyzed publishing strategies.

Our competitions were approved by an international and an
institutional ethics committees. In the competitions, students in a course served as
documents' authors and were assigned to queries.  The students were
incentivized via bonuses to the course grades to write and modify
plain text documents of up to $150$ terms so that the documents will
be highly ranked for the queries. Students in the course could have attained
the perfect grade without participating in the competitions. The
students who participated signed consent forms and could have opted
out at any point in time.

Our bot received a document to be modified
so as to compete with the students for rank promotion. We organized a
competition for each of $15$ queries randomly sampled from the $31$
used by \citet{Raifer+al:17a}\footnote{Raifer et al.'s dataset is
  available at \url{https://github.com/asrcdataset/asrc}.}.  These
queries were originally selected from all topic titles for the Web
tracks of TRECs $2009$-$2012$ by the virtue of having a commercial
intent that can stir up ranking competitions\footnote{The topic IDs of
  the $15$ queries are: $10$, $13$, $18$, $32$, $33$, $34$, $48$, $51$,
  $69$, $167$, $177$, $180$, $182$, $193$, and $195$.}.

  Two students took part in each competition for a query. No two students competed against each other for more than
  one query, and no student competed for more than three queries.  The
  two students competing for a query were provided at the beginning of
  the competition with the same example of a document relevant to the
  TREC topic the query represents.  Some of these documents were
  adopted from the dataset in \citet{Raifer+al:17a}; others were
  created by the authors of this paper using a similar approach to
  that in \citet{Raifer+al:17a}. The students had no incentive to
  stick to these original documents. Hence, in contrast to our bot,
  they had more freedom in promoting their documents. Furthermore, all
  students had prior experience in participating in similar content-based ranking
  competitions for queries other than those they
  were assigned to in our competitions.

  In the first round of the competition, the students submitted their
  modified documents to the search engine. They were then shown a
  ranking induced over a set of five documents: their two documents and
  additional three 
  \firstmention{\planted} documents which were
    randomly selected
    from the first round of
    Raifer et al.'s competitions.
  The students were not aware of the
  fact that other documents might not be written by students 
  competing with them. The identities of
  all documents' authors were anonymized. Throughout the competition,
  the students had access to all documents in rankings.
  The ranking function, described below, was not disclosed to the students.
  
  Having observed the induced ranking, the two students could then
  modify their documents to potentially promote them in the next ranking, and
  submit them for the second round of the competition. The most
  highly ranked \planted document in the first round, which was not
  also the most highly ranked in the entire ranking, and which was
  marked by at least three annotators out of five in Raifer et al's
  competitions as of high quality, was provided
  as input ($\curDoc$) to our bot. Our approach modified the document
  and submitted it ($\newDoc$) to the second round.
  For the other two
  \planted documents, their second-round versions in Raifer et al.'s
  competition were submitted to our second round. As additional baseline which was not part of the actual competitions we use a {\em
  simulated} {\bf static bot}: it receives the same
document in the first round as our bot, and simply submits it to the second round with no
modifications. Comparison with the static bot allows to evaluate
the merit of using a dynamic bot which responds to ranking.

  Our approach was designed for a single shot modification. Hence, our
  main evaluation is based on the ranking induced in the second round
  of the competition with respect to that induced in the first
  round. Furthermore, we used crowdsourcing via the Figure Eight
  platform ({\url{https://www.figure-eight.com}}) to assign
  quality and relevance labels to all documents in the competition as
  in Raifer et al.'s competitions, using their annotation
  guidelines. Each document was annotated by five annotators. A
  document was deemed relevant or of high content quality if it was
  marked as such by at least three annotators. Although not designed
  for iterative document modification, we let our bot participate in
  two additional rounds modifying its document in response to
  rankings. We did not bound the number of previous rankings the bot observes (i.e., the value of~$p$ in Section \ref{sec:features}). The bot utilized in its rank-promotion features information about all rankings as from the first round to the current round. The students had access to the exact same information.

\myparagraph{Document ranking function}
Similarly to \citet{Raifer+al:17a}, we used the state-of-the-art LambdaMART learning-to-rank (LTR) method
\cite{Wu+al:10a} with $25$ standard content-based features as the search engine's ranking function. These features were either used in Microsoft's learning-to-rank datasets\footnote{\url{https://tinyurl.com/rmslr}}, or are query-independent document
quality measures \cite{Bendersky+al:11a}. Further details regarding the ranking method are provided in Appendix \ref{sec:rankFuncApp}.

\myparagraph{Ranking passage pairs}
\label{sec:offline}
Our document modification approach is based on learning-to-rank
passage pairs. As the documents are short (up to $150$ terms), we
used sentences for passages. We trained the approach using the
rankings available for all $31$ queries from round $6$ of Raifer et al.'s competitions; RankSVM was the
passage-pair ranker~\cite{Joachims:06a}. Our training dataset contains $57$ documents which serve for $\curDoc$ and $3399$ passage
pairs $(\psgSrc,\psgTarget)$. Additional details about the
training
are provided in
Appedix \ref{sec:offlineApp}. We now describe the creation of rank-promotion ($\rankPromoteLabel$) and local-coherence maintenance ($\coherenceLabel$), henceforth coherence, labels for training.

For each document $\curDoc$ in a
ranking $\curRank$ in the training set, we create, as described in Section \ref{sec:framework}, passage pairs
$(\psgSrc,\psgTarget)$ where $\psgSrc \in\curDoc$ and
$\psgTarget$ is any passage in documents ranked higher than $\psgSrc$ in $\curRank$. We replace $\psgSrc$ in $\curDoc$
with $\psgTarget$ to yield $\newDoc$, and induce a new ranking
$\nextRank$.\footnote{This procedure is challenging when using APIs of commercial search engines due to the time until the next indexing. We leave this challenge for future work.}
The rank-promotion label, $\rankPromoteLabel$, is $0$ if $\newDoc$'s rank position in $\nextRank$ is the same or worse than that of
$\curDoc$ in $\curRank$; otherwise, $\rankPromoteLabel$ is the
difference between $\newDoc$'s rank in
$\nextRank$ and $\curDoc$'s rank in $\curRank$. As there are $5$
documents in each ranking,
$\rankPromoteLabel$ is in $\set{0,4}$ as mentioned in
Section \ref{sec:framework}.


To produce a coherence label for
$(\psgSrc,\psgTarget)$, we aggregated the labels assigned by human
annotators in two different crowdsourcing tasks performed using
Amazon's Mechanical Turk.
In the first
task,
the annotators were shown $\curDoc$ and $\newDoc$, where
$\psgSrc$ was highlighted in $\curDoc$ and $\psgTarget$ was highlighted
in $\newDoc$.
The annotators were asked to mark which of the two documents was the
original. The coherence label is the number of annotators,
among the $5$ assigned, who failed identifying $\curDoc$
as the original.

The
second coherence label was produced by showing $\newDoc$ to annotators
and telling them that it was obtained by replacing a passage in a document they did not see. The annotators were asked to point to the passage which
presumably replaced a passage in that document; all passages in the document were marked. The number of annotators who did not
identify $\psgTarget$ as the replacing passage is the coherence label.

We scale the arithmetic mean of the two coherence labels by $\frac{4}{5}$ to have
the resultant coherence label, $\coherenceLabel$,
in $\set{0,1,\ldots,4}$ as is the case for the rank-promotion
label. We then use the harmonic mean to aggregate the coherence and rank-promotion labels as described in Section \ref{sec:labels}. The resulting label, $l$, is a real number in $\set{0,4}$. To {\em train} the bot for the
competitions, we
set $\beta=1$ in the harmonic mean; i.e., coherence and rank promotion are
assigned the same weight for training. We demonstrate in the offline
evaluation (Section \ref{sec:offlineResults}) the merits of using this value of $\beta$ with respect to alternatives.

\myparagraph{Offline evaluation}
\label{sec:offlineSetting}
As was the case for training the bot for our competitions (the online
evaluation), we used the round-6 data from \citet{Raifer+al:17a} to
train our approach for offline evaluation. Training is performed with
the labeled ($l$) passage pairs as above. We experiment with $l=c$ and
$l=r$ where only the coherence and rank-promotion labels are used,
respectively; and, with $l$ being the harmonic mean of $c$ and $r$ as
described in Section \ref{sec:labels} with $\beta=1$.\footnote{Note
  that setting $\beta$ to $0$ or to a very high value does not result
  in the integrated labels $l$ relying only on coherence ($c$) or
  rank-promotion ($r$), respectively. Hence, to emphasize these two
  extreme cases, we use $c$ or $r$ as the integrated label $l$.}

To test our approach, we let it modify each of the documents at ranks
$2$--$5$ in round $7$ of Raifer et al.'s \cite{Raifer+al:17a}
competitions for each of the $31$ queries. Each of these documents was
written by a student.  Thus, in total we have $124$ different
experimental settings for an instance of our approach: 4 ranks
$\times$ 31 queries\footnote{In practice we have $113$ different
experimental settings for an instance as we do not
  apply our approach on documents deemed of low quality or keyword
  stuffed in \shortcite{Raifer+al:17a}.}.  Specifically, in each setting,
our approach modifies document $d_{curr}$ into $d_{next}^{bot}$. The
student who originally submitted $d_{curr}$ in round 7 (potentially)
modified it to $d_{next}^{student}$ for round 8. Then, we contrast the rank position of $d_{next}^{bot}$ and $d_{next}^{student}$ in two rankings induced over five documents. Four of the documents are shared between the two rankings: those from round 8 which were submitted by the four students whose documents we did not select for modification in round 7. The fifth document is either $d_{next}^{bot}$ or $d_{next}^{student}$.

In each setting we induce a ranking, using the ranker described above, over the document
modified by our approach and the four round-8 documents of the 
students whose documents were not selected to be modified by our
approach in round 7.\footnote{The ranker we use here is the base ranker Raifer et al. \cite{Raifer+al:17a} utilized. Raifer et al. also added post-hoc rank penalties for low quality documents which we do not apply here. Additional details can be found in Appendix \ref{sec:rankFuncApp}.}  We also induce a ranking over
all five original documents from round 8 of
\citet{Raifer+al:17a}. We then contrast the rank position of
our modified document in the first ranking with that of the student's
modified document in the second ranking. The other four documents are the original student documents from round 8.

We also use five annotators as in the online competitions to evaluate
the quality of documents produced by our approach. Quality annotations
for all other documents are available from Raifer et
al.~\cite{Raifer+al:17a} as described in Appendix \ref{sec:offlineApp}.

To summarize, the offline evaluation is based on contrasting our bot
with a ``human'' agent (student): we let the two modify the same
document which was written by the student. We then contrast the
rank-promotion, quality and relevance of the two modified documents
where all other documents in the same competition for the query at
hand were modified by other students. In contrast to the online
evaluation with our competitions described above, we cannot run this
process for more than a single round. The reason is that the students in Raifer et
al.'s competitions \cite{Raifer+al:17a} did not respond to rankings
that included the documents produced by our approach.



\subsection{Online Evaluation Results}
\label{sec:onlineResults}



There are
five ``players'' per query in each live competition round: two students from our competitions, two students
from Raifer et al.'s competitions whose documents were planted, and
the bot.

We analyzed the competitions using five evaluation measures.
The first three measures quantify ranking properties, and are computed
per player and her document for a query. The values for our students
and Raifer et al.'s are averaged over
the students. The averaged values, and the values for the bot and the static bot, are averaged over
queries. The first measure, {\bf average rank}, is 
the rank of the player's document, averaged as described
above; the highest rank is $1$. The {\bf raw promotion} and {\bf
  scaled promotion} measures quantify the change of a document's rank between rounds $1$ and~$2$. The documents at rank $1$ are not considered as they can only be demoted. Raw promotion is the number of
positions by which the document was promoted (demoted).
The scaled promotion is the raw
promotion normalized by the maximum potential for promotion/demotion with respect to the
document's position. 



The {\bf quality} scores are assigned by crowdsourcing annotators (in
our and in Raifer et al.'s competitions) and attest to the document's
content quality. A document quality score is set to $1$ if at least
three out of five annotators marked it as of high content quality;
otherwise, its quality score is $0$. Using the same approach, we
assigned documents with a $0$/$1$ {\bf relevance} grades for the TREC
topic represented by the query. For the quality and relevance
measures, we report the ratio of queries for which the player's document received a quality/relevance score of 1, and normalize with respect to the number of players where
needed.

\begin{table}[t]
  \caption{\label{tab:main} Online evalution: main result. The best result in a block for each round is boldfaced. Promotion is with respect to the previous round, and hence, there are no promotion numbers for the first round. Recall that positive values for raw and scaled promotion attest to actual promotion while negative values attest to demotion. The lower the values of average rank the better. (The highest rank is 1.)}
    \small
  \center
  \begin{tabular}{@{}llcc@{}}
\toprule
& & round 1 & round 2 \\ \midrule
  \multirow{4}{*}{average rank} & students & $3.200$ & $\;\:3.400$\\
  & \planted & $\mathbf{2.733}$ & $\;\:2.766$\\
  & static bot & $3.133$ & $\;\:3.399$ \\ 
  & our bot & $3.133 $ & $\mathbf{\;\:2.667 }$\\ \midrule
  \multirow{4}{*}{raw promotion} & students & NA & $-0.200$\\
  & \planted & NA & $\;\:0.000$ \\
  & static bot & NA & $-0.266$  \\
  & our bot & NA & $\mathbf{\;\:0.466 }$ \\ \midrule
  \multirow{4}{*}{scaled promotion} & students & NA & $-0.136$ \\
  & \planted  & NA & $-0.002$\\
  & static bot & NA & $-0.177$ \\
  & our bot & NA & $\mathbf{\;\:0.122 }$ \\ \midrule
  \multirow{4}{*}{quality} & students & $0.867$ & $\;\:0.900$\\
  & \planted  & $0.400$ & $\;\:0.766$ \\
  & static bot & $1.000$ & $\mathbf{\;\:1.000}$  \\
  & our bot & $1.000 $ & $\;\:0.933 $\\ \midrule
    \multirow{4}{*}{relevance} & students & $0.933$ & $\;\:0.866$ \\
  & \planted  & $0.900$ & $\mathbf{\;\:0.966}$ \\
  & static bot & $0.866$  &  $\;\:0.866$ \\
  & our bot & $0.866 $ & $\;\:0.866 $ \\ \bottomrule
\end{tabular}

  \end{table}
\myparagraph{Main results}
Table \ref{tab:main} presents our main results for the online evaluation (our competitions).\footnote{We do not present here statistical significance reports as only $15$ queries were used and this is a too small number for computing statistical significance. We do report statistical significance for the offline evaluation in Section \ref{sec:offline} where we use $31$ queries.} Recall that the document
the bot (and the static bot) received per query in the first round was of high quality. The static bot did not change the document for the second round in contrast to our bot.

We see in Table \ref{tab:main} that by all three ranking-based 
evaluation measures, our bot outperformed the two active students in our competitions (``students''), the two students from Raifer et al.'s competitions (``planted'') and the static bot. It is the only player who
has positive raw and scaled promotion values in the second round. Furthermore, 
the bot's documents started from an average rank slightly better than
that of the active students' documents ($3.133$ vs.  $3.2$), and after the modifications for the second round they were
promoted to a higher average rank ($2.667$ vs $3.4$); in fact, on average, the students were demoted from average rank of $3.2$ to $3.4$ which is reflected in the negative raw and scaled promotion numbers. The documents the bot received in the first round were ranked higher than, or the same as,  those of the active students for $53\%$ of the queries. The percentage increased to $67\%$ in the second round after the documents were modified by the bot and the students.

Comparison of our bot with the static bot in terms of average rank and rank promotion attests to the importance of a ``live'' bot
which responds to rankings. We also see that the average quality of
the documents of our bot in the second round ($0.933$) is higher than
that of the documents of the two students from our competitions
($0.9$) and from Raifer et al.'s competitions ($0.766$). The quality
of the static bot's document in the second round is $1$ as it is the
same document from the first round (and the same document our bot
received) which by selection was of quality value $1$.

Table \ref{tab:main} also shows that the document our bot received in
round~$1$ from Raifer et al.'s competitions was not always
relevant. (It was relevant in $86.6\%$ of the cases.) The bot did not
hurt relevance, on average, by its modifications (second round). This
is in contrast to the two active students who had (on average) a lower fraction of relevant documents in the
second round than in the first.

\newcommand{\figWidth}{4.6cm}
\newcommand{\figHeight}{3.6cm}
\begin{figure*}[t]
\hspace*{-0.5cm}
  \begin{tabular}{cccc}
\includegraphics[width=\figWidth, height=\figHeight]{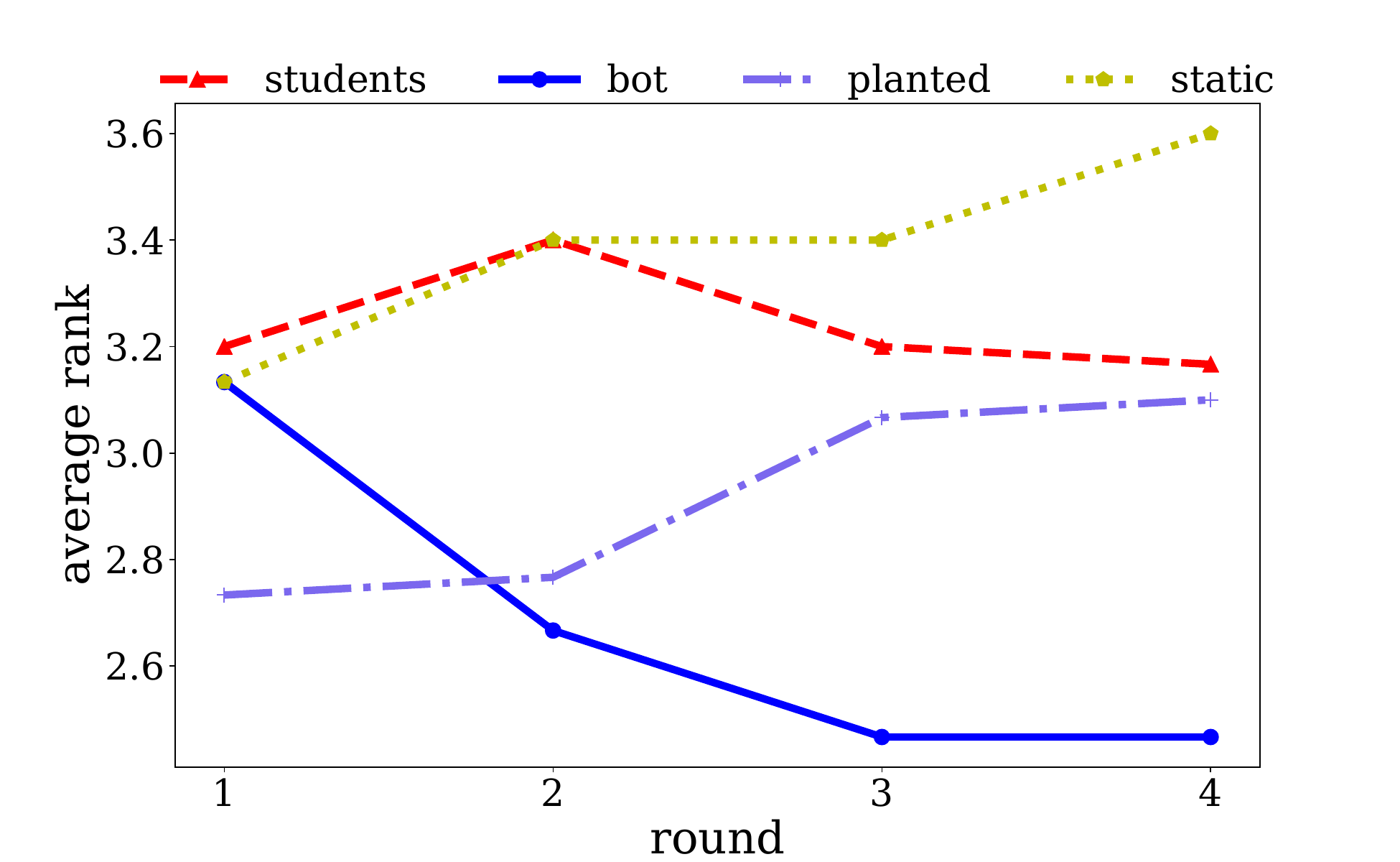} &
\hspace*{-.2in}        {\includegraphics[width=\figWidth, height=\figHeight]{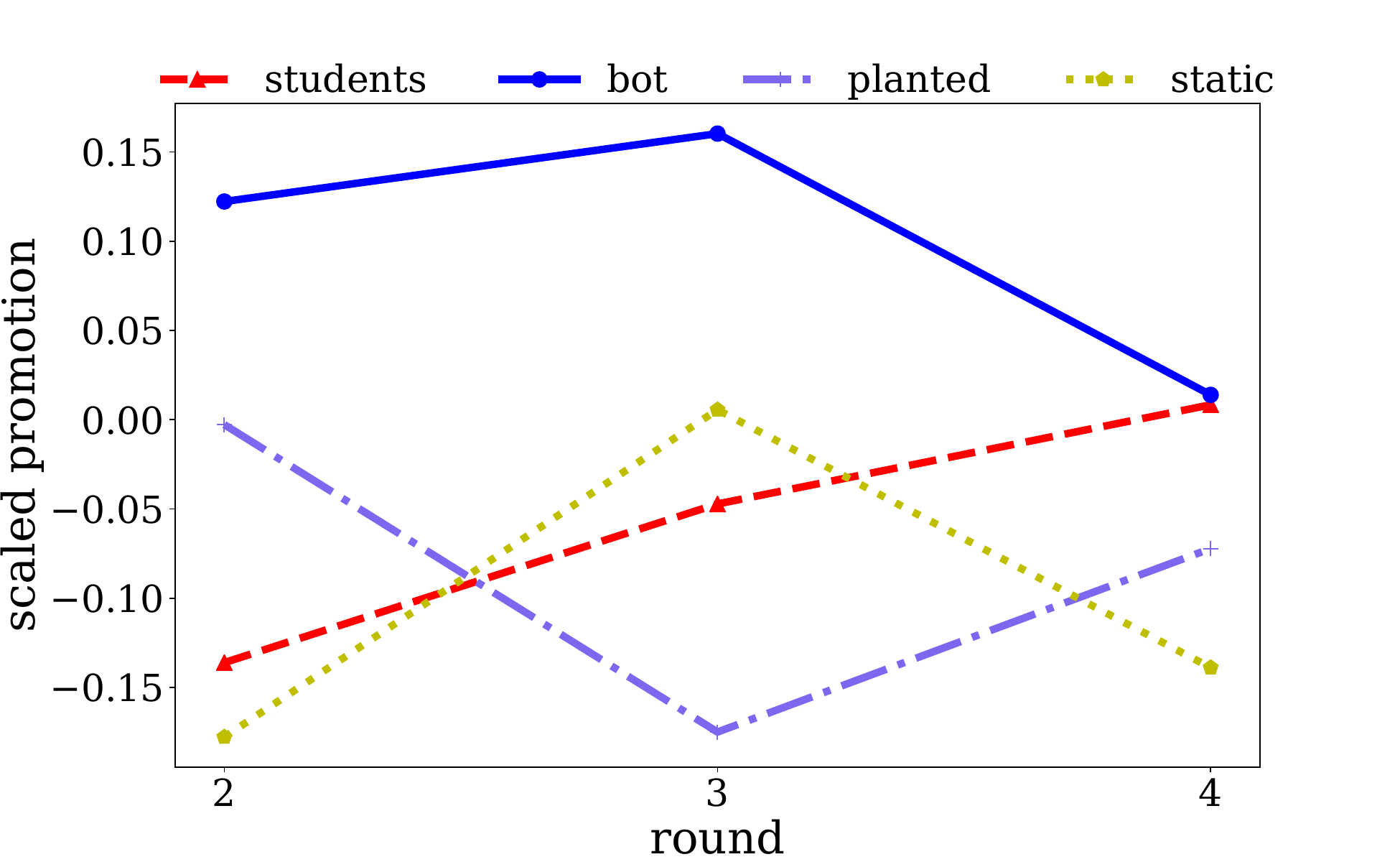}} &
\hspace*{-.2in}  \includegraphics[width=\figWidth, height=\figHeight]{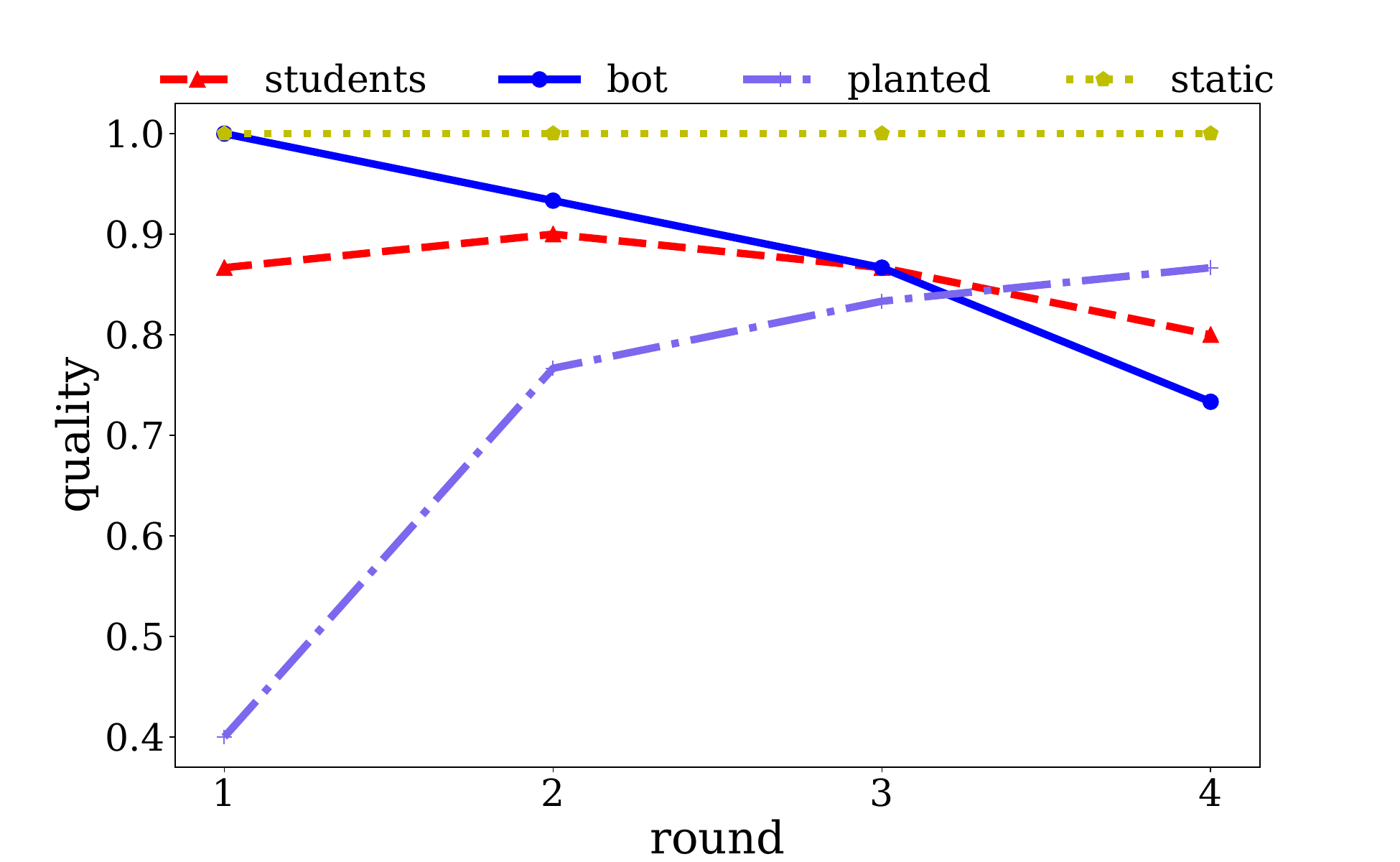} & 
\hspace*{-.2in} \includegraphics[width=\figWidth, height=\figHeight]{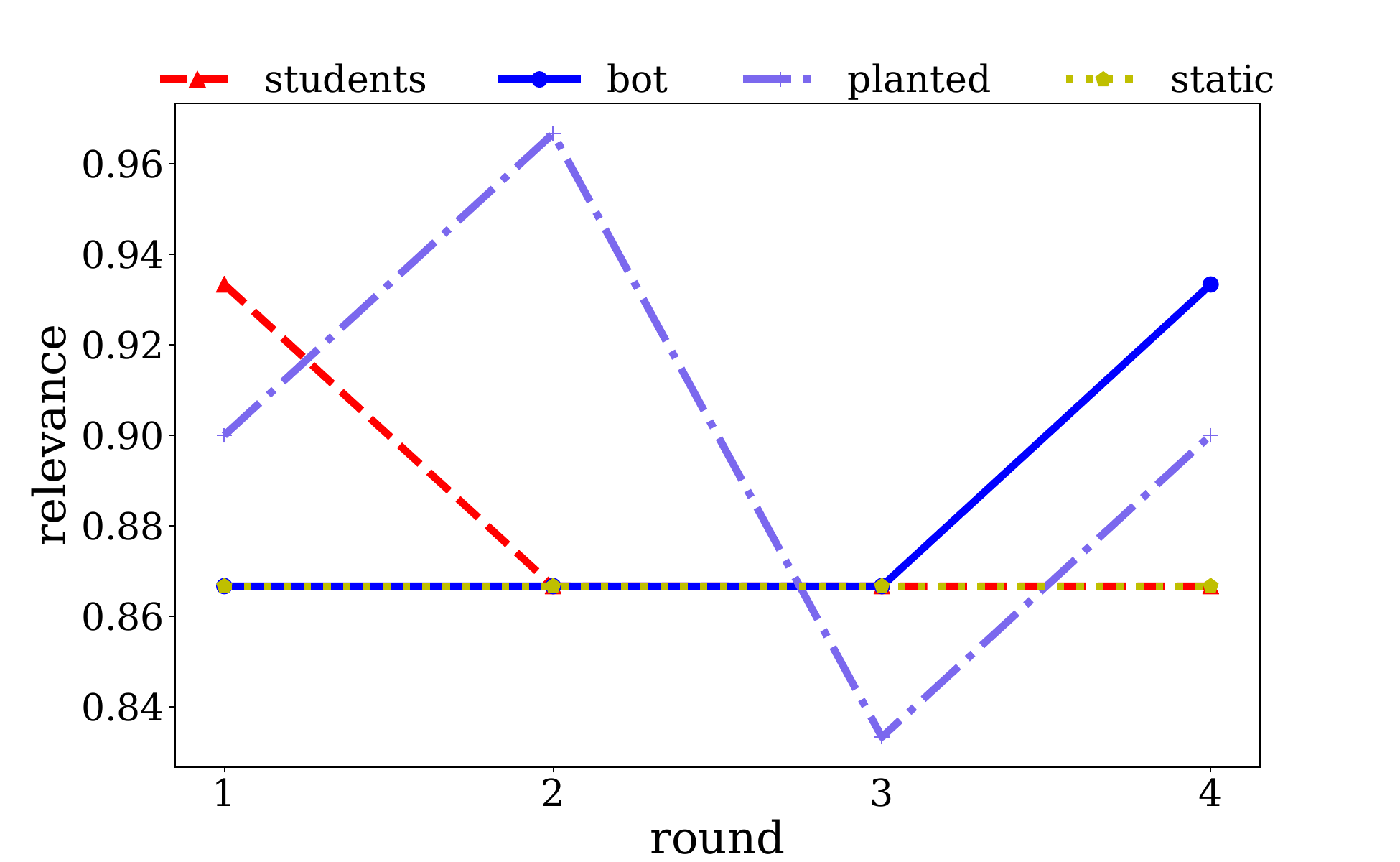} \\
  \end{tabular}
  \caption{\label{fig:competition} Online evaluation: analysis of the four rounds of the
    competition. The curves for the quality and relevance of the
    static bot are horizontal lines with values of $1$ and $0.867$, respectively.}
  \end{figure*}

The competitions' results are encouraging. The bot won
over the active students in terms of rank-promotion and did not hurt relevance in contrast to the students. There was some mild quality decrease as a result of the bot modifications, but the resultant quality still transcends that of the students' documents. In the offline evalution reported in Section \ref{sec:offlineResults}, we show that there is no statistically significant difference between the quality of documents produced by the bot and that of documents produced by students.

Now, while the students had prior experience in ranking
competitions, the bot learned from a past static snapshot of a competition. (See Section \ref{sec:expSet}.)
Moreover, the students could have modified their documents in round $2$ without maintaining faithfulness to their documents from round $1$. This was not the case for the bot by design.


\subsubsection{Repeated Competitions}
Our bot was designed for a single shot (modification) response to a ranking. Yet, we let the competitions run for additional two
rounds with the bot and the two participating students.
The two planted documents from
round~$i$~($\in \set{2,3}$) were replaced with their round $i+1$
versions from Raifer et al.'s competition.
We also use as a baseline the static bot which did not change the
document our bot received in round $1$.

We see in Figure \ref{fig:competition} that in terms of average rank,
the bot wins over all other players and the static bot as from round
2. (This is the first round in which the bot modified the document it
received.) Furthermore, the bot is the only player whose
scaled-promotion values are always non-negative\footnote{The trends
  for raw promotion are similar and hence these results are omitted as they convey
  no additional insight.}. These findings attests to the merits of the
bot in terms of rank promotion.

Figure \ref{fig:competition} also shows that the quality of the bot's
documents monotonically decreases. This is not a surprise as the bot
was designed for a single modification rather than a chain of
modifications; e.g., we did not prevent duplicate sentences in the
modified documents, which the annotators penalized in terms of
quality. Yet, we note that even in round $3$, $85\%$ of the documents
produced by the bot were considered of high content quality by the
annotators; and, in rounds 2 (the first round in which the bot started
changing the document) and 3 the quality of the bot's documents was
higher or equal to that of the documents of the two students who participated in the
competitions. A similar, although less steep, drop of quality is
observed as from round $2$ for the documents produced by the students
who participated in the competition. The increasing quality for the
planted documents can be attributed to the fact that in Raifer et
al.'s competitions there were heavy ranking-penalties for producing
low-quality documents \cite{Raifer+al:17a}, which we did not impose in
our competitions.

Finally, we see in Figure \ref{fig:competition} that the bot did not
cause a decrease in the fraction of relevant documents as a
result of its modifications. In contrast, the average fraction of relevant documents of the students who participated in our
competitions was lower in rounds 2-4 than in round 1.

\setlength{\tabcolsep}{5pt}
\begin{table}[t]
  \caption{\label{tab:offEvaluation} Offline evaluation. Our bot was trained for coherence ($c$), rank-promotion ($r$) and both ($l$); $l$ is the harmonic mean of $c$ and $r$ using $\beta=1$. Statistically significant differences with the students and the static bot are marked with `$s$' and `$b$', respectively. The best result in a column is boldfaced.}
  \small
  \begin{tabular}{@{}llllll@{}}
  \toprule
 & average & raw & scaled & \multirow{2}{*}{quality}  & \multirow{2}{*}{relevance} \\
  & rank & promotion & promotion &  &  \\ \midrule
   students & $3.327$ & $0.212$ &  $0.034$ & $0.991$ & $0.796$\\
static bot & $3.248$ & $0.292$ &  $0.094$ & $\mathbf{1.000}$ & $0.823$\\\midrule
  our bot ($c$) & $3.133$ & $0.407$ &  $0.145$ & $0.973$ & $0.973^{sb}$\\ 
  our bot ($r$) & $\mathbf{2.584^{sb}}$ & $\mathbf{0.956^{sb}}$ &  $\mathbf{0.340^{sb}}$ & $0.973$ & $0.982^{sb}$\\ 
  our bot ($l$) & $2.673^{sb}$ & $0.867^{sb}$ &  $0.309^{sb}$ & $0.982$ & $\mathbf{0.991^{sb}}$\\  \bottomrule
  
\end{tabular}

  \end{table}

\begin{figure*}[t]
\hspace*{-0.5cm}
  \begin{tabular}{cccc}
    \includegraphics[width=\figWidth, height=\figHeight]{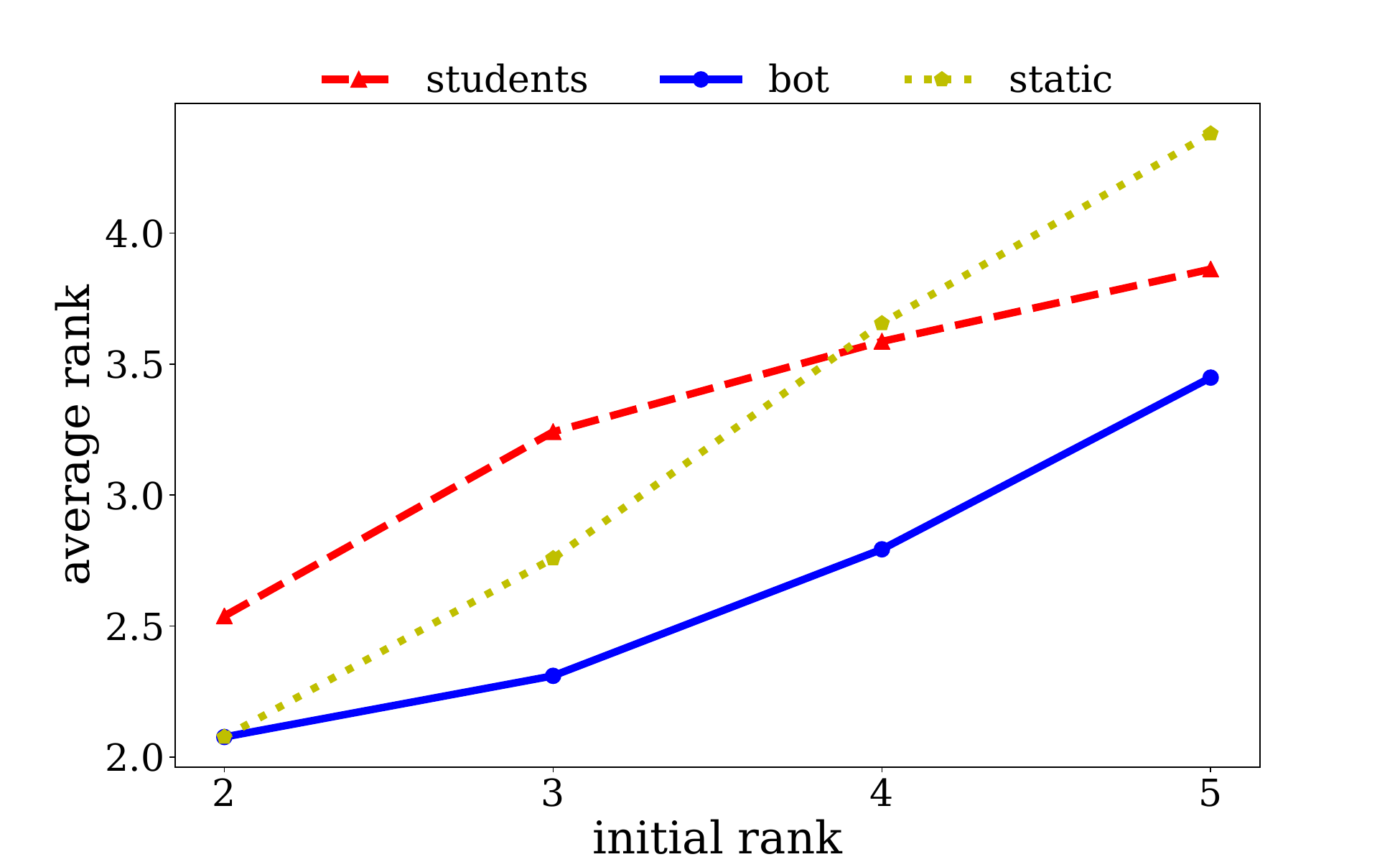} &
    \includegraphics[width=\figWidth, height=\figHeight]{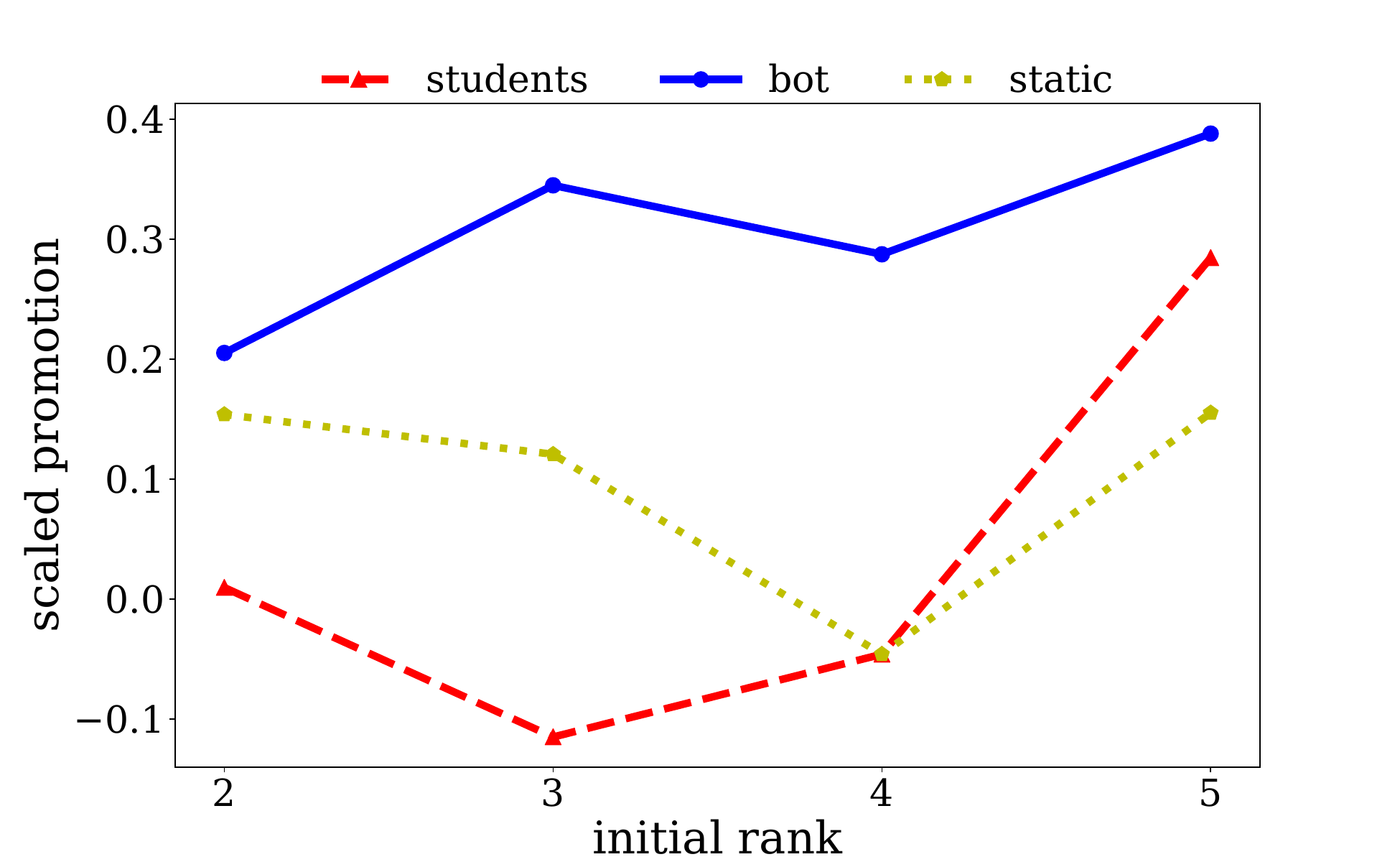} &
    \hspace*{-0.5cm}  \includegraphics[width=\figWidth, height=\figHeight]{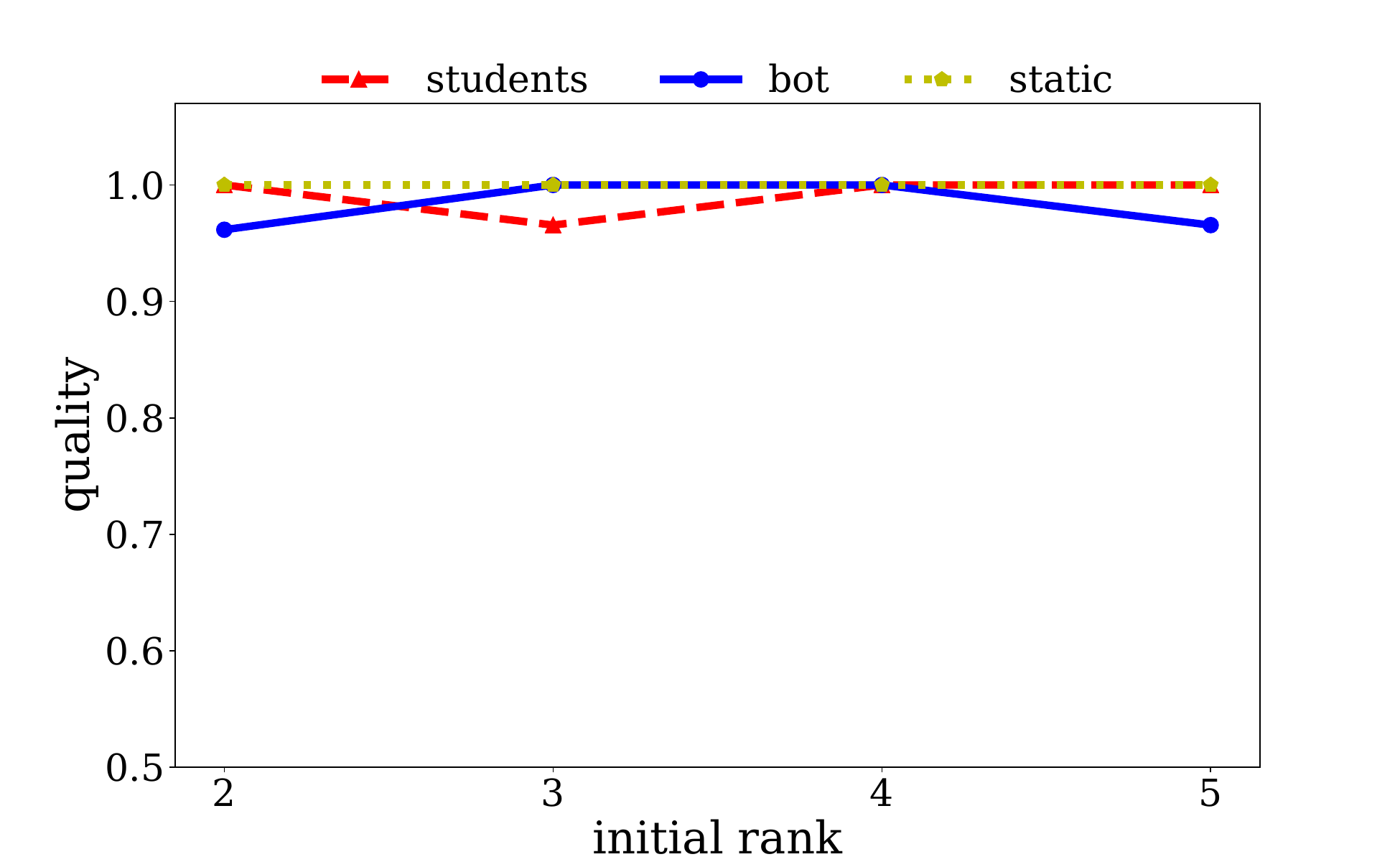} & 
    \hspace*{-0.5cm}  \includegraphics[width=\figWidth, height=\figHeight]{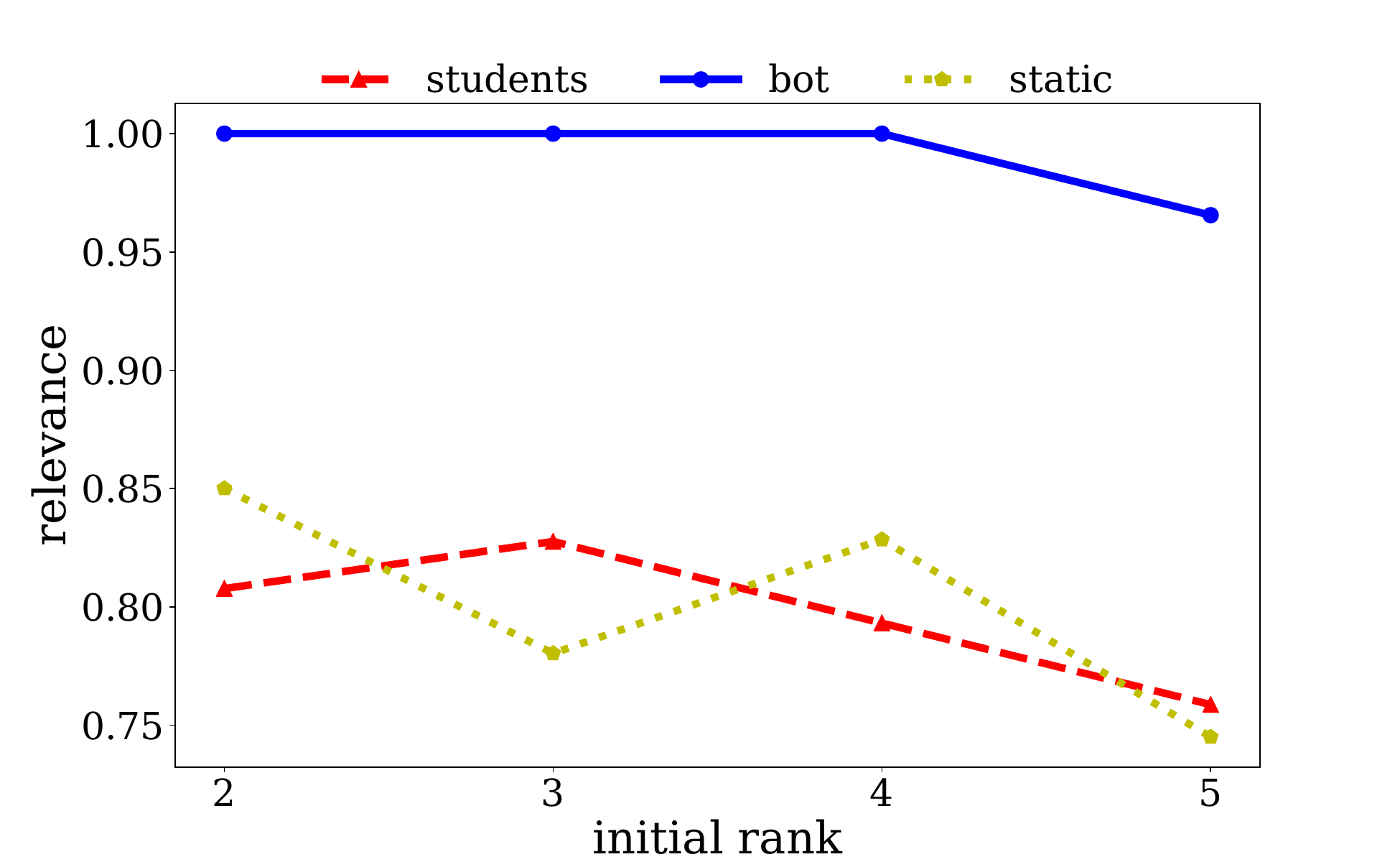} 
\end{tabular}
  \caption{\label{fig:offline} Offline evaluation: our bot (trained with
    the $l$ labels and $\beta=1$) vs. the student(s). Both modify the same document at
    the same initial rank for a query. For reference comparison we use a static bot that receives the same document and does not modify it. The presented numbers are averages over $31$ queries. Recall that lower rank means
    higher positioning.}
  \end{figure*}

\subsection{Offline Evaluation Results}
\label{sec:offlineResults}
We now turn to describe the offline evaluation results. Recall from
Section \ref{sec:offlineSetting}, which describes the experimental
setting, that the evaluation is performed using the competitions of
Raifer et al. \cite{Raifer+al:17a}. Specifically,
our approach, henceforth referred to as bot, is applied to documents in round 7.
The approach
is trained with three types
of labels which results in three bots: one trained as in the online evaluation
for both rank-promotion and coherence ($l$ labels) with $\beta=1$ in
the harmonic mean; the other two are trained either only for coherence
($c$ labels) or only for rank-promotion ($r$ labels).



Table \ref{tab:offEvaluation} presents the average over (initial) ranks
($2$--$5$) and 31 queries of the rank-promotion, quality and relevance
measures for the documents produced by our three bots
and for the documents  produced by the
corresponding students. As in the online evaluation,
we use for reference comparison a static bot which keeps the student
document as is. Figure \ref{fig:offline} presents the
per initial rank measures when training the bot for both coherence and rank-promotion~($l$) with $\beta=1$ as was the case in the online evaluation\footnote{The raw promotion graph is omitted as
  it conveys no additional insight: it shows the exact same
  patterns as in the scaled promotion graph.}.
Statistically significant differences for all measures are computed
(over the $31$ queries) using the two tailed permutation
(randomization) test (with $100000$ random permutations) at a 95\% confidence level. Bonferroni correction
was applied for multiple testing.

We see in Table \ref{fig:offline} that all three versions of our bot
outperform the students and the static bot for all three
rank-promotion measures: average rank, raw promotion and scaled
promotion. The improvements are substantial and statistically significant when using the $r$ (only rank promotion) and $l$ (rank promotion
and coherence) labels for training the bot. Furthermore, the fraction of relevant documents produced by
each of the three bot versions is statistically significantly higher
than that for the students and the static bot. The fraction of quality documents produced by the three bot versions is slightly lower than that of the students, but the differences are never statistically significant.

Among the three versions of our bot, the one trained for both coherence and rank-promotion yields the highest quality and relevance results and posts very strong rank-promotion performance (the second best in the table); hence, it was selected for the online evaluation discussed above. We further see in Table \ref{tab:offEvaluation} that, as expected, training for rank-promotion --- alone ($r$) or together with coherence ($l$) --- results in much better rank promotion than when training only for coherence ($c$). Furthermore, all the differences in average rank, raw promotion and scaled promotion between using $r$ and $c$ and between using $l$ and $c$ are statistically significant. The quality of the produced documents does not vary much with respect to the version used for training; indeed, none of the quality differences between the three versions is statistically significant. In terms of relevance, training for both rank-promotion and coherence~($l$) outperforms training only for promotion ($r$) and training only for coherence ($c$); however, none of the differences between the three, in terms of relevance performance, is statistically significant.

The averages over initial ranks reported in Table \ref{fig:offline} for the $l$-label bot well
reflects the per initial rank state-of-affairs shown in Figure~\ref{fig:offline}: the documents produced by our bot are (i) more
highly ranked, (ii) of quality that is statistically indistinguishable from, and~(iii) more often relevant with respect to the students' documents. 
These findings are also aligned with those presented in Section~\ref{sec:onlineResults} for the online evaluation. All in all, both the online and offline evaluations attest to the clear merits of our proposed approach.



\subsection{Feature Analysis}
Table \ref{tab:features} presents the feature weights learned by the
RankSVM passage-pair ranker which was used in the online and offline evaluations: training was performed with rank-promotion and coherence integrated labels ($l$) with $\beta=1$ in the harmonic mean. Appendix~\ref{sec:offlineApp} provides additional details of training the RankSVM.  Feature weights are comparable as feature
values are min-max normalized.

We see in Table \ref{tab:features} that the weight of the \qryTermTarget feature,
which is a measure of query-terms occurrences in the passage to be
used for replacing another, is the highest. Indeed, using passages that
contain many occurrences of query terms can help to improve retrieval
scores and hence ranking. In accordance, the feature with the lowest negative weight is \qryTermSrc which quantifies the query-terms occurrences in the passage that is candidate for being replaced. Indeed, the more query terms it contains, the less likely its replacement is to promote the document in a ranking.

The next three features with the highest
weights are \simTargetTopTF, \simTargetPrevTopWtV and \simTargetTopWtV. These are
measures of the (lexical and semantic) similarity of a candidate replacing passage to the
documents most highly ranked in the current (\simTargetTopTF, \simTargetTopWtV) and
previous (\simTargetPrevTopWtV) rankings. This finding 
provides further support to the merits of mimicking documents most highly
ranked in the past.

Other features with positive weights include the semantic
similarity of the candidate replacing passage with the passage to be replaced
(SimSrcTarget(W)) and its preceding passage
in the document (SimTargetPrecPsg(W)). These features quantify the potential change of local coherence as a result of the passage replacement. 


\begin{table}[t]
  \scriptsize
  \caption{\label{tab:features} Feature weights of the passage-pair ranker.}
  \center
  \begin{tabular}{@{}lc@{}}
\toprule
Feature & Weight \\
\midrule
QueryTermTarget & $\;\:\:0.189$ \\ 
SimTargetTop(TF.IDF) & $\;\:\:0.134$ \\ 
SimTargetPrevTop(W2V) & $\;\:\:0.138$ \\
SimTargetTop(W2V) & $\;\:\:0.085$ \\ 
SimSrcPrevTop(W2V) & $0.084$ \\ 
SimTargetPrecPsg(W2V) & $0.034$ \\ 
SimSrcPrecPsg(W2V) & $\;\:\:0.024$ \\ 
SimSrcTarget(W2V) & $0.015$ \\ 
SimTargetFollowPsg(W2V) & $\;\:\:0.015$ \\ 
SimSrcTop(W2V) & $\;\:\:-0.013$ \\ 
SimSrcFollowPsg(W2V) & $-0.015$ \\ 
SimSrcPrevTop(TF.IDF) & $\;\:\:-0.020$ \\ 
SimTargetPrevTop(TF.IDF) & $\;\:\:-0.022$ \\ 
SimSrcTop(TF.IDF) & $\;\:\:-0.025$ \\ 
QryTermSrc & $-0.073$ \\ 
\bottomrule
\end{tabular}

\end{table}



\section{Conclusions}
We presented a novel method of modifying a document so as to promote
it in rankings induced by a non-disclosed ranking function for a given
query. The only information about the function is past rankings it
induced for the query. Our method is designed to maintain the content
quality of the document it modifies.

Our method 
replaces a passage of the document with another passage --- a challenge we address as a learning-to-rank task over
passage pairs with a dual-objective: rank promotion and content-quality (coherence) maintenance.

Our method served as a bot in content-based ranking competitions
between students. The bot produced documents that were of high
quality, and better promoted in rankings than the
students' documents. The bot's modifications did not hurt relevance in contrast to the modifications introduced by students. Additional offline evaluation further demonstrated the merits of our bot.

\paragraph*{Ethical considerations}
Worries about potential abuse of our method for black hat SEO can be
alleviated: the method is tuned for maintaining content
quality. Furthermore, as the ranking competitions show, the method's
potential negative effects on the search echo system are not significant, and can be smaller than
those introduced by human authors who try to promote documents. Dropping the
constraint of quality maintenance in our method will result in the produced
documents being of low quality. But in this case, simple quality estimates used in Web
search methods \cite{Bendersky+al:11a} can be used to easily disqualify these documents or penalize them in rankings.

\myparagraph{Acknowledgments}
We thank the reviewers for their comments.
The work by Moshe Tennenholtz and Gregory Goren was supported by funding from the European Research Council (ERC) under the European Union’s Horizon 2020 research
and innovation programme (grant agreement 740435).

\appendix
\section{Appendix}
\label{sec:app}
We next provide some additional technical details about the
experimental setting.
\subsection{Document Ranking Function}
\label{sec:rankFuncApp}
For document ranking, we used the same learning-to-rank approach, and features, used by \citet{Raifer+al:17a} in their fundamental ranker. Specifically,
the state-of-the-art LambdaMART learning-to-rank (LTR) approach
\cite{Wu+al:10a}\footnote{We used the RankLib implementation: \url{www.lemurproject.org/ranklib.ph}.}  was used with $25$ content-based features. (Recall that
we focus on content-based modifications.) These features were either
used in Microsoft's learning-to-rank datasets\footnote{\url{https://tinyurl.com/rmslr}}, or are query-independent document
quality measures --- specifically, stopword-based measures and the
entropy of the term distribution in a document --- demonstrated to be
effective for Web retrieval \cite{Bendersky+al:11a}.

To train the ranking function, we used the ClueWeb09 Category~B collection
and its $200$ topic-title queries (TREC $2009$-$2012$). We used the query likelihood retrieval approach \cite{Song+Croft:99a} with Dirichlet
smoothed document language models; the smoothing parameter was set to
$1000$~\cite{Zhai+Lafferty:01a}. Documents assigned a score below $50$
by Waterloo's spam classifier were removed from rankings. The resultant top-$1000$ documents were used for training. We used default
values of the free parameters in the implementation except for the number of leaves and trees
which were selected from $\set{5,10,25,50}$ and $\set{250,500}$,
respectively, using five-fold cross validation performed over queries:
four folds were used for training and one for validation of these two
parameters; NDCG@5 was the optimization criterion. We select the
parameter values that result in the best NDCG@5 over all $200$ queries
when these were used as part of the validation folds.

\subsection{Learning-To-Rank Passage Pairs}
\label{sec:offlineApp}
Our approach is based on ranking for a given document $\curDoc$, which we want to rank-promote, passage pairs
$(\psgSrc,\psgTarget)$, where $\psgSrc$ is a passage in $\curDoc$ and $\psgTarget$ is a
passage in a document among the most highly ranked in the current
ranking, $\curRank$. (Refer back to Section \ref{sec:framework} for
details.)  We use a learning-to-rank method to learn a passage-pair ranker and to apply it. To train our approach, we used all $31$ queries and all the
documents submitted for these queries in round $6$ of Raifer et al.'s
competition \cite{Raifer+al:17a}\footnote{In round $5$ of Raifer et al.'s
  competition, the incentive system has changed
  \cite{Raifer+al:17a}. Hence, we selected a round which is after the
  change.}, except for those which were not marked as of high quality
by at least $3$ out of~$5$ crowdsourcing annotators
\cite{Raifer+al:17a}. To induce document ranking, we used the 
ranking function from Appendix \ref{sec:rankFuncApp} which was also used in the ranking competitions. Recall that our approach has no knowledge of the document ranking
function.

We let our approach modify a document, $\curDoc$, which is either the
lowest ranked or the second highest ranked 
for a query. Thus, we have a mix of low ranked and high ranked
documents which we let our approach train with. As a result, for each
query we consider two identical current rankings, $\curRank$, over
the given documents. In each of these two rankings, a single document
--- ranked second or last --- is designated as $\curDoc$. And, we
induce two new rankings, $\nextRank$, where $\curDoc$ was modified by
our approach to $\newDoc$. The rest of the documents are not modified;
i.e., we train our approach by assuming that other documents do
not change\footnote{The alternative would have been to have other
  documents modified simultaneously to $\curDoc$. However, this would have introduced much noise
  to the learning phase as the ranking of $\newDoc$ could have
  changed with respect to that of $\curDoc$ not necessarily due to the modification of $\curDoc$, but rather
  due to those of others.}. Our training dataset contains $57$ documents which serve for $\curDoc$ and $3399$ passage pairs $(\psgSrc,\psgTarget)$. For each document there are, on average, $59.6$ such pairs (standard deviation: $42.32$) to be ranked.

Some of the features on which our approach relies, namely
{\simSrcPrevTopTF}, {\simTargetPrevTopTF}, {\simSrcPrevTopWtV} and
{\simTargetPrevTopWtV}, utilize information about the past $p$
rankings. For training, we let
our approach observe all current and past rankings (i.e., $p=6$) where these were
induced using our ranking function over the documents in each of the
first five rounds of Raifer et al.'s competition \cite{Raifer+al:17a}. Recall from Section \ref{sec:expSet} that we also do not bound~$p$ when we apply the bot in the online evaluation.

As noted in Section \ref{sec:framework}, any feature-based
learning-to-rank approach can serve for our passage-pair ranking
function. Since we do not have large amounts of training data, we used
a linear RankSVM~\cite{Joachims:06a}\footnote{\url{https://www.cs.cornell.edu/people/tj/svm_light/svm_rank.html}};
all free parameters were set to default values of the implementation,
except for $C$ which was set using cross validation. Specifically, we
used $5$ fold cross validation over all $57$ documents\footnote{Recall that ranking of passage pairs is with respect to a specific document $\curDoc$.} which serve as $\curDoc$, where $4$ folds were used to train RankSVM and one fold (validation) was used to
set $C$'s value ($\in \set{0.001, 0.01, 0.1}$) by optimizing for
NDCG@5. NDCG is computed for the ranking of passage pairs with their assigned $l$ labels; these lables are also used for training; see Section \ref{sec:expSet} for details. As each document is part of a single validation fold, we set $C$ to the value that optimized NDCG@5 over all documents when these were part of a validation fold. We then trained the approach with this $C$ value using all documents and used it as a bot in the online and offline evaluations.

As described in Section \ref{sec:framework}, our approach utilizes, as
features, Word2Vec-based similarities. We train a query-based Word2Vec
model \cite{Diaz+al:16a}, so as to rely on the query context, using
the {\rm gensim}
package (\url{https://radimrehurek.com/gensim/models/word2vec.html}). Specifically,
for a query $\query$, the model was trained on the top-$10000$
documents retrieved from ClueWeb09 Category B for $\query$ using the
query likelihood retrieval model \cite{Song+Croft:99a}; spam removal
was not applied here. Default parameter values of the {\rm gensim}
package were used, except for the threshold of number of occurrences
per word which was set to $0$, the window size which was set to $8$,
and the vector size which was set to $300$.


\balance
\bibliographystyle{ACM-Reference-Format}
\bibliography{cunlp-ir,coherence}
\end{document}